\documentclass[fleqn,twoside,12pt]{article}
\usepackage{latexsym,amssymb} 

\def\p{\partial}

\begin{document}

\begin{flushleft}
\Large \bf Equivalence of {\mathversion{bold}$Q$}-Conditional Symmetries \\
under Group of Local Transformation
\end{flushleft}

\noindent
{\bf Roman O. POPOVYCH}

\bigskip

\noindent
{\small\it Institute of  Mathematics of
the  National  Ukrainian Academy of Sciences, \\
3 Tereshchenkivska Str. 3, 01601 Kyiv,  Ukraine}

\bigskip

\noindent  {\small URL:
{\tt http://www.imath.kiev.ua/\~{}rop/} \\
E-mail: {\tt rop@imath.kiev.ua}}

\begin{abstract}
\noindent
The definition of $Q$-conditional symmetry for one  PDE is correctly generalized
to a special case of systems of PDEs and involutive families of operators. 
The notion of equivalence of  $Q$-conditional symmetries under a group of 
local transformation is introduced. Using this notion,  
all possible single $Q$-conditional symmetry operators are classified 
for the \mbox{$n$-dimensional} ($n\geqslant 2$) linear heat equation and 
for the Euler equations describing the motion of an incompressible ideal fluid.
\end{abstract}

\noindent
The concept of $Q$-conditional symmetry called also nonclassical symmetry 
was introduced by Bluman and Cole in 1969. This year is the year of the $30^{\rm th}$ 
anniversary of appearance of their pioneering paper \cite{roman:bluman&cole}. 
Although the concept of $Q$-conditional symmetry exists for a long time and has 
various applications many problem of its theory are not solved so far.

Before 1986 the nonclassical symmetry was only mentioned in  few papers, e.g., 
in \cite{roman:harrison&estabrook}. The intensive application of $Q$-conditional symmetries 
to finding exact solutions of partial differential equations (PDEs) and the parallel search for 
their foundations was begun after publication of the papers of  Olver and Rosenau 
in~1986 and~1987 \cite{roman:olver&rosenau86,roman:olver&rosenau87} as well as the paper of 
Fushchych and Tsyfra in 1987 \cite{roman:fushchych&tsyfra87}.

The first correct definition of a $Q$-conditional symmetry operator for one  PDE was 
given in~\cite{roman:fushchych&tsyfra87}. Later it was generalized to involutive families of 
operators [6--8].
We stress that it can be directly extended only to some special cases 
of $Q$-conditional invariance for systems of PDEs. For all the other cases this 
definition must be essentially modified and is much more complicated.

In this paper we correctly generalize the definition of $Q$-conditional symmetry 
[6--8]
to a special case of systems of PDEs and involutive families of operators. 
Further, we introduce the notion of equivalence of  $Q$-conditional symmetries 
under a group of local transformation. Using this notion,  we can, first, 
classify all the possible  $Q$-conditional symmetries and, correspondingly, 
all the possible reductions of systems of PDEs 
\cite{roman:zhdanov&tsyfra96,roman:zhdanov&tsyfra&popovych99} 
and, secondly, essentially simplify the procedure of finding 
$Q$-conditional symmetries in some cases when 
the Lie symmetry group is sufficiently wide.

Consider a system of $k$ PDEs of the order $r$ 
for $m$ unknown functions $u=(u^1,\ldots,u^m)$ 
depending on $n$ independent variables $x=(x_1,\ldots,x_n)$
of the form
\begin{equation}\label{roman:general.system.of.PDEs}
L(x,u_{(r)}(x))=0, \qquad L=(L^1,\ldots,L^k).
\end{equation}
Here the order of a system is 
the order of the major partial derivative 
appearing in the system. The symbol $u_{(r)}$ denotes 
for the set of partial derivatives of the functions $u$ 
of the orders from 0 to $r$. Within the local approach 
system~(\ref{roman:general.system.of.PDEs}) is treated 
as a system of algebraic equations in the jet space $J^{(r)}$ 
of the order $r$. 

Consider also an involutive family Q of $l$ differential operators 
\begin{equation}\label{roman:general.involutive.family.of.operators}
Q^s=\xi^{si}(x,u)\p_{x_i}+\eta^{sa}(x,u)\p_{u_a}, 
\qquad \mbox{where} \qquad l\leqslant n, \quad 
\mathop{\rm rank}\nolimits ||\xi^{si}(x,u)||=l.
\end{equation}
The requirement of involution  means for the family~$Q$ 
that the commutator of any pair of operators from~$Q$ belongs to 
the span of $Q$ over the ring  of smooth functions of the variables~$x$ 
and $u$, i.e. 
\begin{equation}\label{roman:involution.condition} 
\forall \, s,p\quad  \exists \, \zeta^{sps'}=\zeta^{sps'}(x,u)\mbox{:}\quad 
[Q^s,Q^p]=\zeta^{sps'}Q^{s'}.
\end{equation}
Here and below the indices $a$ and $b$ run from 1 to $m$, 
the indices $i$ and $j$ run from 1 to $n$,  
the indices $s$ and $p$ run from 1 to $l$, and 
the indices $\mu$ and $\nu$ run from 1 to $n-l$. 
The sumation is imposed over the repeated indices.
Subscripts of functions denote dif\/ferentiation with respect to the 
corresponding variables.

If operators~(\ref{roman:general.involutive.family.of.operators}) 
form an involutive family, then the family $\widetilde Q$ 
of differential operators
\begin{equation}\label{roman:equivalent.family.of.operators}
\widetilde Q^s=\lambda^{sp}Q^p, 
\qquad \mbox{where} 
\qquad \lambda^{sp}=\lambda^{sp}(x,u), 
\quad \det||\lambda^{sp}||\not=0,
\end{equation}
is also involutive. And family~(\ref{roman:equivalent.family.of.operators}) 
is called {\em equivalent} to 
family~(\ref{roman:general.involutive.family.of.operators})
[6--8].

\vspace{1ex}

\noindent 
{\bf Notation:} $\widetilde Q=\{\widetilde Q^s\}\sim Q=\{Q^s\}$.

\vspace{1ex}

By the Frobenius theorem, condition~(\ref{roman:involution.condition}) is 
sufficient for the system of PDEs 
\begin{equation}\label{roman:Q.system}
Q^s[u^a]:=\eta^{sa}(x,u)-\xi^{si}(x,u)\frac{\p u^a}{\p x_i}=0
\end{equation}
to be compatible. 

Denote the manifold defined by the system of algebraic equations 
$L=0$ in $J^{(r)}$ by $\cal L$ and the manifold corresponding to 
the set of all the differential consequences of the system of 
PDEs~(\ref{roman:Q.system}) in $J^{(r)}$ by $\cal M$:
\[\arraycolsep=0pt\begin{array}{ll}
{\cal L}&=\{ (x,u_{(r)}) \in J^{(r)}\, |\, L(x,u_{(r)})=0\},\\[1ex]
{\cal M}&=\{ (x,u_{(r)}) \in J^{(r)}\, |\, D_1^{\alpha_1}\ldots D_n^{\alpha_n}Q^s[u^a]=0, 
\;\: \alpha_i\in\mathbb{N}\cup\{0\},\;\:|\alpha|\mbox{:}=\alpha_1+\cdots+\alpha_n<r \},
\end{array}\] 
where $ D_i=\p_{x_i}+\sum\limits_{\alpha}u^a_{\alpha,i}\p_{u^a_\alpha}$
is the operator of total differentiation with respect to the variable~$x_i$, 
$u^a_\alpha$ and $u^a_{\alpha,i}$ denote the variables in $J^{(r)}$, corresponding to 
derivatives $ \frac{\p^{|\alpha|}u}{\p x_1^{\alpha_1}\ldots\p x_n^{\alpha_n}}$
and $ \frac{\p^{|\alpha|+1}u}{\p x_1^{\alpha_1}\ldots
\p x_{i-1}^{\alpha_{i-1}}\p x_i^{\alpha_i+1}\p x_{i+1}^{\alpha_{i+1}}\ldots\p x_n^{\alpha_n}}$.

\vspace{1ex}

Let the system~$L|_{\cal M}=0$ do not includes equations 
which are differential consequences of other its equations. 
Moreover, let all the differential consequences of the system~$L|_{\cal M}=0$, 
the orders of which (as equations)  are less than or equal to its order, vanish on ${\cal L}\cap{\cal M}$.

\vspace{1ex}

\noindent
{\bf Definition 1.} {\it System of smaller PDEs~(\ref{roman:general.system.of.PDEs}) is called 
{\em $Q$-conditional invariant} with respect to involutive family 
of differential operators~(\ref{roman:general.involutive.family.of.operators}) 
if the relation
\begin{equation}\label{roman:definition.of.Q-conditional.symmetry}
\Bigr(Q^s_{(r)}L\Bigl)\Bigl|_{\;{\cal M}\cap{\cal L}}\:=0
\end{equation}
holds true. Here the symbol $Q^s_{(r)}$ denotes the $r$th prolongation 
of the operator $Q^s$:
\[
Q^s_{(r)}=Q^s+\sum_{|\alpha|{}\leqslant  r} \eta^{sa\alpha}\p_{u^a_\alpha}, 
\qquad
\eta^{sa\alpha}=D_1^{\alpha_1}\ldots D_n^{\alpha_n}
(\eta^{sa}-\xi^{si}u^a_i)+\xi^{si}u^a_{\alpha,i}.
\]}

Denote the set of involutive families of $l$ operators 
of $Q$-conditional symmetry of 
system~(\ref{roman:general.system.of.PDEs}) as ${\cal B}({\cal L},l)$: 
\[
{\cal B}({\cal L},l)=\left\{ Q=\{Q^1,\ldots,Q^l\} \,\left|\: 
\begin{array}{l}\mbox{the system $L=0$ is  $Q$-conditionally}
\\ \mbox{invariant with respect to }Q \end{array}\right.\right\}
\]

\vspace{1ex}

\noindent
{\bf Lemma [6--8].} {\it Let system of PDEs~(\ref{roman:general.system.of.PDEs}) be 
$Q$-conditionally invariant with respect to involutive family of 
operators~(\ref{roman:general.involutive.family.of.operators}). 
Then, it is $Q$-conditionally invariant with respect to 
an arbitrary family of the form~(\ref{roman:equivalent.family.of.operators}), i.e. 
\[
Q\in {\cal B}({\cal L},l),\; \widetilde Q\sim Q\quad \Longrightarrow\quad  
\widetilde Q\in {\cal B}({\cal L},l).
\]}

An important consequence of the lemma is that we can study 
$Q$-conditionally invariance up to equivalence 
relation~(\ref{roman:equivalent.family.of.operators}) 
which is defined on the set of involutive families of $l$~ope\-ra\-tors
as well as in ${\cal B}({\cal L},l)$. Then it is possible for an arbitrary family 
of operators~(\ref{roman:general.involutive.family.of.operators}) 
to choose the functions $\lambda^{sp}(x,u)$ and, if it is necessary, 
to change enumeration of the variables~$x_1$,~\dots,~$x_n$ in such a way 
that operators~(\ref{roman:equivalent.family.of.operators}) take the following form: 
$\widehat Q^s=\p_{x_s}+\hat\xi^{s,l+\nu}\p_{x_{l+\nu}}+\hat\eta^{sa}\p_{u^a}.$
Operators $\widehat Q^s$ generate a commutative Lie algebra. 

Let $A({\cal L})$ and $G({\cal L})$ denote the maximal Lie invariance 
algebra of system~(\ref{roman:general.system.of.PDEs}) and its maximal
local symmetry correspondingly. Now we strengthen the equivalence 
relation in ${\cal B}({\cal L},l)$, given by formula~(\ref{roman:equivalent.family.of.operators}), 
by means of generalizing equivalence of $l$-dimensional subalgebras 
of the algebra~$A({\cal L})$ under the adjoint representation of 
the group~$G({\cal L})$  in $A({\cal L})$.

We use the following lemma for this generalization.
 
\vspace{1ex}

\noindent
{\bf Lemma.} {\it Let $g$ be an arbitrary local transformation from $G({\cal L})$. 
Then the adjoint action of $g$ in the set of differential operators generate 
a one-to-one mapping from ${\cal B}({\cal L},l)$ into itself.}

\vspace{1ex}

Let $Q=\{Q^s\}$ and $\widetilde Q=\{\widetilde Q^s\}$ be involutive families of 
differential operators.
 
\vspace{1ex}

\noindent
{\bf Definition.} {\it The families $Q$ and $\widetilde Q$ are called equivalent 
with respect to a group $G$ of local trans\-formations if there exists a local 
transformation $g$ from $G$ for which the families $Q$ and 
${\rm Ad}(g)\widetilde Q$ are equivalent.}
 
\vspace{1ex}

\noindent
{\bf Notation:} $Q\sim \widetilde Q \bmod G.$

\vspace{1ex}

\noindent
{\bf Definition.} {\it The families $Q$ and $\widetilde Q$ are called equivalent 
with respect to a Lie algebra $A$ of dif\-fe\-rential operators if they are equivalent
with respect to the one-parametric group generated by an operator from $A$.}
 
\vspace{1ex}

\noindent
{\bf Notation:} $Q\sim \widetilde Q \bmod A.$

\vspace{1ex}

{\advance\topsep-4pt
Therefore,
\begin{equation}\label{roman:group.equiv.relation}
Q\sim \widetilde Q \bmod G\quad \stackrel{\rm def}{\Longleftrightarrow} \quad 
\exists \, g\!\in\! G\mbox{:}\; Q\sim {\rm Ad}(g)\widetilde Q.
\end{equation}
\begin{equation}\label{roman:algebra.equiv.relation}
Q\sim \widetilde Q \bmod A\quad \stackrel{\rm def}{\Longleftrightarrow} \quad 
\exists \, V\!\in\! A\mbox{:}\; Q\sim\widetilde Q \bmod 
\{e^{\varepsilon V},\; \varepsilon\!\in\! U(0,\delta)\subset {\mathbb R}\}
\end{equation}
}

\noindent
{\bf Lemma.} {\it Formulas~(\ref{roman:group.equiv.relation}) and~(\ref{roman:algebra.equiv.relation}) define 
equivalence relations in the set of involutive families of $l$~dif\-ferential operators. 
Moreover, if $G$ is a subgroup of $G({\cal L})$ and $A$ is a subalgebra of $A({\cal L})$ then 
formulas~(\ref{roman:group.equiv.relation}) and~(\ref{roman:algebra.equiv.relation}) define 
equivalence relations in ${\cal B}({\cal L},l)$.}

\vspace{1ex}

Comsider two examples.

\vspace{1ex}

\noindent
{\bf Example 1.} Investigate $Q$-conditional invariance 
of the linear $n$-dimensional heat equation 
\begin{equation}\label{roman:lhe}
u_t=u_{aa}, \quad \mbox{where} \quad u=u(t,\vec x), \quad 
t=x_0, \quad \vec x=(x_1,\ldots,x_n), 
\end{equation}
with respect to a single operator ($l=1$).

It is just the problem with $n=1$ for which Bluman and Cole 
introduced the concept of nonclassical symmetry. 
In the one-dimensional case the problem was completely 
solved in \cite{roman:fshsp}. That is why we pay our 
attention to  the multidimensional problem.

Lie symmetry of equation~(\ref{roman:lhe}) is well known.
In the one-dimensional case it was investigated by Lie. 
The maximal Lie invariance algebra~$A(\mbox{LHE})$ of~(\ref{roman:lhe})
is generated by the following operators:
\begin{equation}\label{roman:alhe}\arraycolsep=0pt
\begin{array}{l} 
\p_t=\p /\p t, \quad 
\p_a=\p /\p x_a, \quad 
D=2t\p_t+x_a\p_a, \quad 
G_a=t\p_a-\frac{1}{2}x_au\p_u, \quad 
I=u\p_u, \\[1ex] 
J_{ab}=x_a\p_b-x_b\p_a \; (a<b), \quad 
\Pi=4t^2+4tx_a\p_a-(x_ax_a+2t)u\p_u, \quad 
f(t,\vec x)\p_u, 
\end{array}\end{equation}
where $f=f(t,\vec x)$ is an arbitrary solution of~(\ref{roman:lhe}). 

\vspace{1.5ex}

\noindent
{\bf Theorem 1.} {\it For any operator $Q$ of $Q$-conditional symmetry of 
equation~(\ref{roman:lhe}) one of three following conditions holds:
\begin{enumerate}
\item 
$Q\sim \widetilde Q^0,$ \quad 
where \quad $\widetilde Q^0\!\in\! A\mbox{\rm (LHE)};$
\item
$Q\sim \widetilde Q^1=\partial_n+g_n g^{-1}u\partial_u  
\bmod ASO(n)+A^\infty\mbox{\rm (LHE)},$ 
where $g=g(t,x_n)$ ($g_n\not=0$) is a solution of the one-dimensional 
heat equation, that is, $g_t=g_{nn}$;
\item
$Q\sim \widetilde Q^2=J_{12}+\varphi(\theta) u\partial_u  
\bmod AG(1,n)+A^\infty\mbox{\rm (LHE)},$
where $\varphi=\varphi(\theta)$ is a solution of the equation
$\varphi_{\theta\theta}+2\varphi\varphi_\theta=0,$ 
$\varphi_\theta\not=0,$ $\theta$~is the polar angle in the plane $OX_1X_2.$
\end{enumerate}
Here 
\[
\arraycolsep=0pt\begin{array}{l}
A^\infty\mbox{(LHE)}=\langle f(t,\vec x)\partial_u|f=f(t,\vec x)\mbox{:}f_t=f_{aa}\rangle,
\vspace{1mm}\\
AG(1,n)=\langle\p_t,\p_a,G_a,J_{ab}\rangle, \qquad 
ASO(n)=\langle J_{ab}\rangle.
\end{array}
\]}

It follows from Theorem 1 that there exist only three classes 
of the possible reductions on one independent variable for 
the linear multidimensional heat equation. 

The first class is formed by Lie reductions.

The second class involves reductions which are similar to separation of variables 
in the Cartesian coordinates:
\[
u=g(t,x_n)v(\omega_0,\ldots,\omega_{n-1}), \quad \mbox{where} \quad 
\omega_0=t,\; \omega_i=x_i;
\qquad (\ref{roman:lhe})\quad \Longrightarrow \quad v_0=v_{ii}.
\]

The third class is formed by reductions which are similar to separation of variables 
in the cylindrical coordinates:
\[\arraycolsep=0pt
\begin{array}{l} 
u=\exp\bigr(\int\!\varphi(\theta)d\theta\bigr)v(\omega_0,\ldots,\omega_{n-1}), 
\quad\mbox{where} \quad 
\omega_0=t,\; \omega_1=r,\; \omega_s=x_{s+1},\: s=\overline{2,n-1}; 
\\[1ex] (\ref{roman:lhe})\quad \Longrightarrow \quad 
v_0=v_{11}+\omega_1^{-1}v_1-\lambda\omega_1^{-2}v+v_{ss}.
\end{array}\]
Here $\lambda=-\varphi_\theta-\varphi^2=\mathop{\rm const}\nolimits,$
$(r,\theta)$~are the polar coordinates in the plane $OX_1X_2.$
As the equation $\varphi_{\theta\theta}+2\varphi\varphi_\theta=0$ 
has four essentially different (under translations with respect to $\theta$) 
families of solutions with $\varphi_\theta\not=0,$
there are four inequivalent cases for the third class of reductions 
$(\varkappa\not=0)$:
\[\arraycolsep=0pt\begin{array}{lll}
\mbox{a) }\: \varphi=-\varkappa\tan \varkappa\theta:\quad
&u=v(\omega_0,\ldots,\omega_{n-1})\cos\varkappa\theta,\quad 
&\lambda=\varkappa^2;
\\[1ex]
\mbox{b) }\: \varphi=\varkappa\tanh \varkappa\theta:\quad
&u=v(\omega_0,\ldots,\omega_{n-1})\cosh \varkappa\theta,
\quad 
&\lambda=-\varkappa^2;
\\[1ex]
\mbox{c) }\: \varphi=\varkappa\coth \varkappa\theta:\quad
&u=v(\omega_0,\ldots,\omega_{n-1})\sinh \varkappa\theta,
\quad 
&\lambda=-\varkappa^2;
\\[1ex]
\mbox{d) }\: \varphi=\theta^{-1}:\quad 
&u=v(\omega_0,\ldots,\omega_{n-1})\theta,\quad 
&\lambda=0.
\end{array}\]

\noindent
{\bf Example 2.} Consider the Euler equations 
\begin{equation}\label{roman:ees}
\vec u_t+(\vec u\cdot\nabla)\vec u+\nabla p=\vec 0, 
\qquad
\mathop{\rm div}\nolimits\vec u=0
\end{equation}
describing the motion of an incompressible ideal fluid.
In the following
$\vec u=\{u^a(t,\vec x)\}$ denotes the velocity of the fluid,
$p=p(t,\vec x)$ denotes the pressure,
$n=3$, $\vec x=\{x_a\},$
$\p_t=\p/\p t,$ 
$\p_a=\p/\p x_a,$ 
$\vec\nabla=\{\p_a\},$
$\triangle=\vec \nabla\cdot\vec \nabla$ is the Laplacian.
The fluid density is set equal to unity.

Lie symmetry of system~(\ref{roman:ees}) was investigated by 
Buchnev~\cite{roman:buchnev,roman:olver}.
The maximal Lie invariance algebra~$A(\mbox{E})$ of~(\ref{roman:ees})
is infinite dimensional and generated by the following operators:
\begin{equation}\label{roman:ae}
\arraycolsep=0pt\begin{array}{l}
\partial_t, 
\quad 
J_{ab}=x_a\p_b-x_b\p_a+u^a\p_{u^b}-u^b\p_{u^a} \quad (a<b),
\\[1ex]
D^t=t\p_t-u^a\p_{u^a}-2p\p_p, \quad D^x=x_a\p_a+u^a\p_{u^a}+2p\p_p,
\\[1ex]
R(\vec m)=m^a(t)\p_a+m^a_t(t)\p_{u^a}-m^a_{tt}(t)x_a\p_p,
\quad 
Z(\chi)=\chi(t)\p_p,
\end{array}
\end{equation}
where $m^a=m^a(t)$ and $\chi=\chi(t)$ are arbitrary smooth functions of 
$t$ (for example, from $C^{\infty}((t_0,t_1),{\mathbb R})$). 
Let us investigate $Q$-conditional symmetry of~(\ref{roman:ees}) with 
respect to alone differential operator 
$Q=\xi^0(t,\vec x,\vec u,p)\p_t+\xi^a(t,\vec x,\vec u,p)\p_a+
\eta^a(t,\vec x,\vec u,p)\p_{u^a}+\eta^0(t,\vec x,\vec u,p)\p_p.$

\vspace{1ex}

\noindent
{\bf Theorem 2.} {\it Any operator $Q$ of $Q$-conditional symmetry of the Euler
equations~(\ref{roman:ees}) either is equivalent to a Lie symmetry operator of~(\ref{roman:ees})
or is equivalent ($\bmod A(\mbox{E})$) to the operator
\begin{equation}\label{roman:nonlie.operator.ees}
\widetilde Q=\p_3+\zeta\left(t,x_3,u^3\right)\p_{u^3}+\chi(t)x_3\p_p, 
\end{equation}
where $\zeta_{u^3}\not=0$,  $\zeta_3+\zeta\zeta_{u^3}=0$, 
$\zeta_t+\left(u^3\zeta+\chi x_3\right)\zeta_{u^3}+(\zeta)^2+\chi=0$.}

\vspace{1ex}

It follows from Theorem 2 that there exist two classes of the possible reductions 
w.r.t. independent variable for the Euler equations, namely, the Lie reductions 
and the reductions corresponding to operators of form~(\ref{roman:nonlie.operator.ees}).

Lie reductions of the Euler equations~(\ref{roman:ees}) are investigated in [12--14].

An ansatz constructed with the operator $\widetilde Q$ has the following form:
\[
u^1=v^1,\quad u^2=v^2,\quad u^3=x_3v^3+\psi\left(t,v^3\right),\quad 
p=q+{\textstyle \frac{1}{2}}\chi(t)x_3^2,
\] 
where $v^a=v^a(t,x_1,x_2)$, $q=q(t,x_1,x_2)$,
$\psi=\psi(t,v^3)$ is a solution of the equation 
\[\psi_t-\left(\left(v^3\right)^2+\chi\right)\psi_{v^3}+v^3\psi=0.\]

Substituting this ansatz into~(\ref{roman:ees}), we obtain the corresponding reduced system 
($i,j=1,2$):
\[
v^i_t+v^jv^i_j+q_i=0,\quad v^3_t+v^jv^3_j+\left(v^3\right)^2+\chi=0,\quad v^j_j+v^3=0.
\]

The analogous problem for the Navier--Stokes equations 
\begin{equation}\label{roman:nses}
\vec u_t+(\vec u\cdot\nabla)\vec u+\nabla p-\nu\triangle\vec u=\vec 0, 
\qquad
\mathop{\rm div}\nolimits\vec u=0 \qquad (\nu\not=0)
\end{equation}
describing the motion of an incompressible viscous fluid 
was solved by Ludlow, Clarkson, and Bassom in \cite{roman:ludlow&clarkson&bassom}.
Their result can be reformulated as follows.

\vspace{1ex}

\noindent
{\bf Theorem 3.} {\it Any (real) operator $Q$ of $Q$-conditional symmetry of the Navier--Stokes 
equations~(\ref{roman:nses}) is equivalent to a Lie symmetry operator of~(\ref{roman:nses}).}

\vspace{1ex}

Therefore, all the possible reductions of the Navier--Stokes equations w.r.t. independent 
variable are exhausted by the Lie reductions.
Lie symmetry of system~(\ref{roman:nses}) was studied by Danilov \cite{roman:danilov,roman:bytev.a}. 
The maximal Lie invariance algebra of the Navier-Stokes equations~(\ref{roman:nses} is similar to 
one of the Euler equations (see~(\ref{roman:ae})):
\[
A({\rm NS})=\langle\p_t,\; J_{ab},\; D^t+{\textstyle\frac{1}{2}}D^x,\; 
R(\vec m(t)),\; Z(\zeta(t))\rangle.
\]
The Lie reductions of the Navier--Stokes equations were completely described in 
\cite{roman:jnmp94}.


\begin{thebibliography}{99}


\bibitem{roman:bluman&cole} Bluman~G. and Cole~J.D., The general similarity
solution of the heat equation, {\em J.~Math. Mech.,} 1969, V.18, N~11, 
1025--1042. 


\bibitem{roman:harrison&estabrook} Harrison~B.K. and Estabrook~F.B., Geometric 
Approach to Invariance Groups and Solutions of Partial Differential Systems, 
{\em J.~Math.~Phys.}, 1969, V.12, N~4 , 653--666. 


\bibitem{roman:olver&rosenau86} Olver~P.J. and Rosenau~P., The construction of special
solutions to partial differential equations, {\em Phys.~Lett.~A}, 1986,
V.114, N~3, 107--112. 

\bibitem{roman:olver&rosenau87} Olver~P.J. and Rosenau~P., Group-invariant solutions 
of differential equations, {\em SIAM J.~Appl.~Math.}, 1987,
V.47, N~2, 263--278. 


\bibitem{roman:fushchych&tsyfra87} Fushchych~W.I. and Tsyfra~I.M., On a reduction and
solutions of nonlinear wave equations with broken symmetry, 
{\it J. Phys. A: Math. Gen.}, 1987, V.20, N~2, L45--L48.


\bibitem{roman:fushchych&zhdanov92} Fushchych~W.I. and Zhdanov~R.Z., 
Conditional symmetry and reduction of partial dif\/ferential equations, 
{\it Ukr. Math. J.}, 1992, V.44, N~7, 970--982.


\bibitem{roman:zhdanov&tsyfra96} Zhdanov~R.Z. and Tsyfra~I.M., 
Reduction of dif\/ferential equations and conditional symmetry, 
{\it Ukr. Math.~J.}, 1996, V.48, N~6, 595--602. 


\bibitem{roman:zhdanov&tsyfra&popovych99}
Zhdanov~R.Z., Tsyfra~I.M. and Popovych~R.O.,
A precise definition of reduction of partial dif\/ferential equations,
{\it J.~Math.~Anal.~Appl.}, 1999, V.238, N~1, 101--123.


\bibitem{roman:fshsp}
Fushchych~W.I., Shtelen~W.M., Serov~M.I. and Popowych~R.O., 
$Q$-con\-di\-tio\-nal symmetry of the linear heat 
equation, {\it Dopov. Acad. Sci. Ukraine.}, 1992, N~12, 27--32.


\bibitem{roman:buchnev} 
Buchnev A.A., The Lie group admitted by the equations of the motion 
of an ideal incompressible fluid, {\it Dinamika sploshnoy sredy}, 
Novosibirsk, Institute of Hydrodynamics, 1971, V.7, 212--214.


\bibitem{roman:olver} Olver P.J., Applications of Lie Groups to
Differential Equations,  New York, Sprin\-ger, 1986.


\bibitem{roman:our.conf.95.halya}
Popovych H., On reduction of the Euler equations by means of 
two-dimensional subalgebras, 
{\it J. Nonlin. Math. Phys.}, 1996, V.3, N~3--4, 441--446.


\bibitem{roman:dopov96.8.halya}
Popovych H., Reduction of the Euler equations to systems in three 
independent variables, 
{\it Dopov. Acad. Sci. Ukraine}, 1996, N~8, 23--29.

\bibitem{roman:akpr}
Andreev~V.K., Kaptsov~O.V., Pukhnachev~V.V. and Rodionov~A.A.,
Application of Group Theoretical Methods in Hydrodynamics,
Novosibirsk, Nauka, 1994. 


\bibitem{roman:ludlow&clarkson&bassom}
Ludlow D.K., Clarkson P.A. and Bassom A.P.,
Nonclassical symmetry reductions of the three-dimensional 
incompressible Navier--Stokes equations, 
{\it J. Phys. A: Math. Gen.}, 1998, V.31, 7965--7980.


\bibitem{roman:danilov}
Danilov Yu.A., 
Group properties of the Maxwell and Navier--Stokes equations, 
Preprint, Acad. Sci. USSR, Kurchatov Institute of Atomic Energy, Moscow, 1967.


\bibitem{roman:bytev.a}
Bytev V.O., Group properties of the Navier--Stokes equations, 
{\it Chislennye metody mekhaniki sploshnoy sredy},  
Comp. Center, Siberian Dep. Acad. Sci. USSR, Novosibirsk, 
1972, V.3, N~4, 13--17. 


\bibitem{roman:jnmp94}
Fushchych W.I. and Popowych R.O.,
Symmetry reduction and exact solution of the Navier--Stokes equations, 
{\it J. Nonlin. Math. Phys.}, 1994, V.1, N~1,~2, 75--113, 158--188.



\end{thebibliography}
\end{document}